\documentclass[a4paper,12pt]{article}

\usepackage[centertags]{amsmath}
\usepackage{amsfonts}
\usepackage{amssymb}
\usepackage{amsthm}
\usepackage{newlfont}
\usepackage[english]{babel}
\usepackage{enumerate}

\newtheorem{theorem}{Theorem}

\theoremstyle{remark}

\theoremstyle{definition}

\theoremstyle{definition}

     \newcommand {\beq}  {\begin{equation}}
      \newcommand {\eeq}  {\end{equation}}  

\newcommand{\bsn}{{\boldsymbol n}}

\author{Aleksenko A.I., J.P.Cruz, Lakshtanov E.L.,  \thanks{Department of
Mathematics, Aveiro University, Aveiro 3810, Portugal.  This work
was supported by {\it Centre for Research on Optimization and
Control} (CEOC) from the ''{\it Funda\c{c}\~{a}o para a
Ci\^{e}ncia e a Tecnologia}'' (FCT), cofinanced by the European
Community Fund FEDER/POCTI, and by the FCT research project
PTDC/MAT/72840/2006.} \thanks{e-mail: lakshtanov@rambler.ru}}

\title{High frequency limit of the Transport Cross Section and boundedness of  the
Total Cross Section in scattering by an obstacle with impedance boundary conditions}

\begin{document}
\date{}
\maketitle

\begin{abstract}
The scalar scattering of the plane wave by a strictly convex
obstacle with impedance boundary conditions is considered.  The
uniform boundedness of the Total Cross Section for all values of
frequencies is proved. The high frequency limit of the Transport
Cross Section is founded and presented as a classical functional
of the variational theory.
\end{abstract}

\section{Introduction}



Consider a strictly convex body $\Omega \subset \mathbb R^3$ with
Lipschitz boundary $\partial \Omega$ and $k > 0$. The scattered
field is given by the Helmholtz equation and a radiation condition
\begin{equation}\label{helm}
\Delta u(r)+k^2 u(r)=0, \quad r \in \Omega'=\mathbb R^3 \backslash
\Omega
\end{equation}
\begin{equation}\label{Somm}
\int_{|r|=R} \left |\frac{\partial u(r)}{\partial n}-iku(r) \right
|^2 dS = o(1), \quad R \rightarrow \infty
\end{equation}
with Dirichlet, Neumann or impedance boundary conditions of the
form
\begin{equation}\label{dir}
\mathcal B_\gamma (u)|_{\partial \Omega} \equiv - \mathcal
B_\gamma (e^{ik(r \cdot \theta_0)})|_{\partial \Omega} , \quad
\quad r=(x,y,z) \in
\partial \Omega, \quad
\end{equation}
where  $\gamma \geq 0$ is a constant, $\mathcal B_\gamma
=(\partial /
\partial n)-ik\gamma$ and $e^{ik(r \cdot \theta_0)}$ is an incident field formed by
a plane wave with incident angle $\theta_0=(0,0,1) \in S^2$. In \cite{vantychonovsamarsky}, for example, is proved the existence and
uniqueness of the solution of (\ref{helm})-(\ref{dir}). A function
$u(r)$ which satisfies the mentioned conditions has asymptotic
\begin{equation}\label{scamp}
u(r)=\frac{e^{ik|r|}}{|r|} f_\gamma(\theta)+o \left
(\frac{1}{|r|} \right ), \quad r \rightarrow \infty, \quad
\theta=r/|r| \in S^2
\end{equation}
where function $f_\gamma(\theta)=f_\gamma(\theta,k)$ is called
{\it scattering amplitude} and measure
$$
\sigma_\gamma=\int_{S^2} |f_\gamma(\theta)|^2 d\sigma(\theta)
$$
is called Total Cross Section. $\sigma$ is a square element of the
unit sphere. The projection on the incident direction $\theta_0$
of the total momentum transmitted to the obstacle is given by
a measure called Transport Cross Section (for a large volume
normalization)
\begin{equation}\label{rrr}
 \quad R_\gamma=\int_{S^2} <\theta_0-\theta,\theta_0>|f_\gamma(\theta)|^2 d\sigma(\theta).
\end{equation}

In case of Dirichlet (or Neumann\footnote{To simplify expressions
we will use only notation of $f_0(\theta)$ in formulas
(\ref{uuo})-(\ref{rrl}), but all of them are valid also for the
case $\gamma=\infty$. See \cite{majdaTay} for proof of $\lim_{k
\rightarrow \infty} \sigma_\infty = 2\sigma_{cl}$}) boundary
conditions the limit behavior of the scattering amplitude in the
high frequency regime is described completely by following two
statements: \cite[Theorem 1]{majda}
\begin{equation}\label{uuo}
f_0(\theta)=\frac{1}{2}\mathcal
K(y^+(\theta))^{-1/2}e^{ik<y^+(\theta)\cdot (\theta -
\theta_0)>}+O(1
 / k), \quad \theta \neq \theta_0.
\end{equation}
Here $y^+(\theta) \in \partial \Omega$ is the preimage of
$\bsn(\theta):=(\theta-\theta_0)/(|\theta-\theta_0|) \in S^2$
under the Gauss map,
$\mathcal K(y^+)$ is the Gauss curvature at $y^+ \in \partial
\Omega$. The estimation $O(1/k)$  is uniform on compact subsets of
$\{ \theta \in S^2 \,|\, \theta \neq \theta_0 \}$.

The behavior near the forward directions is given by (see
\cite{sph,wl})
\begin{equation}\label{adl}
\lim_{k \rightarrow \infty} |f_0|^2=|f|^2_{cl}+\sigma_{cl}
\delta(\theta_0)
\end{equation}
where $|f_{cl}(\theta)|^2=(2K(y^+(\theta)))^{-1}$ is the classical
density of scattered rays, and the limit is in sense of distributions.
 This formula allows one to
obtain limits at high $k$ of all measures like
\begin{equation}\label{phie}
\int_{S^2} \varphi(\theta) |f_0(\theta)|^2d\sigma(\theta), \quad
\varphi \in C(S^2).
\end{equation}
 In particularly we have,
\begin{equation}\label{rrl}
\lim_{k \rightarrow 0} \sigma_0=2\sigma_{cl}, \quad \lim_{k
\rightarrow 0} R_{0}=R_{cl},
\end{equation}
Here $\sigma_{cl},R_{cl}$ are classical Total Cross Section and
classical Transport  Cross Section. Calculation of the limit of
the $\sigma_0$ in the sphere case is mentioned in every student
book of physics, and for the case of convex bodies there is a
rigorous proof in \cite{majdaTay}. Moreover, this fact was used in
\cite{sph} to prove (\ref{adl}).

The case of impedance boundary condition (i.e. finite values of
$\gamma >0$) is not completely studied. Exists an analog of
(\ref{uuo}) (\cite[Th.1.]{majda}) namely,
\begin{equation}\label{uuogamma}
f_\gamma(\theta)=\frac{1}{2}\mathcal
K(y^+)^{-1/2}e^{ik<y^+(\theta)\cdot (\theta - \theta_0)>}\left (
\frac{\gamma - <\bsn(\theta),\theta>} {\gamma + <\bsn(\theta),
\theta>} \right )+O(1
 / k)
\end{equation}
uniformly on every open subset of $\{ \theta \in S^2 : \theta \neq \theta_0\}$ for $k
\rightarrow \infty$.
But unfortunately the behavior of $\sigma_\gamma$ for $\gamma>0$ and
large values of $k$ is not investigated. Behavior of the
scattering amplitude near the forward direction $\theta_0$ is
unknown and therefore we don't have tools to calculate limits of
measures like (\ref{phie}), even if density $\varphi$ turns
to zero at $\theta_0=0$, like it happens with Transport Cross
section (\ref{rrr}). But it becomes possible if we will prove that
$\sigma_\gamma$ is bounded from infinity uniformly for all (large
enough) values of $k$.

\begin{theorem}\label{maj}
1. Exists a number $C=C(\gamma)>0$ such that for all $k \geq 0$
\begin{equation}\label{majest}
\sigma_\gamma \leq C.
\end{equation}

2. Let the visible part of $\Omega$ be written
as a graph of the smooth function $g(x) : \mathcal I \rightarrow
\mathbb R^3$, where $\mathcal I \in \mathbb R^2$ is a part of
plane perpendicular to $\theta_0$.  Then
\begin{equation}\label{eqmaj}
\lim_{k \rightarrow \infty}  R_\gamma = \int_{\mathcal I} \frac{2
dx}{1+|\nabla g|^2} \left ( \frac{\gamma \sqrt{1+|\nabla g|^2}
-1}{\gamma \sqrt{1+|\nabla g|^2} + 1} \right )^2.
\end{equation}
\end{theorem}
Note that this integral has the form of a standard functional of the
optimization theory. Cases $\gamma=0$ and $\gamma = \infty$
correspond to the classical resistance functional which was
investigated starting from Newton (1686) (for example
\cite{Principia}-\cite{CL1}) and in many recent articles. So, we
suppose, that extremal problems of the constructed functional
(\ref{eqmaj}) is an object for investigation.
\subsection{Proofs}

The proof of Theorem~\ref{maj} is based on the results of
\cite{majda}.  Let's prove the first part of the theorem \ref{maj}:

Proof. Since $f(\theta,k)$ is a continuous function on $k \geq 0$
(see \cite{AlRamm}) we should prove that (\ref{majest}) holds for
large values of $k$.

In \cite[Prop. 2.1, p. 273]{majda} is proved that scattering
amplitude has estimation
\begin{equation}\label{majaux1}
|f_\gamma(\theta)-\widetilde{f}_\gamma(\theta)| \leq
\frac{C_m(\gamma)}{k^m}, \quad \theta \in S^2
\end{equation}
where $\{C_m(\gamma)\}$ are positive constants and $m$ can be
chosen arbitrary large. Function $\widetilde{f}(\theta)$ is
determined from the expression
$$
\widetilde{u}(r)=(e^{ik|r|}/|r|)(\widetilde{f}(\theta)+o(1)),
\quad r \rightarrow \infty.
$$
In own turn  field $\widetilde{u}(r)$ is determined through
\begin{equation}\label{majaux2}
\widetilde{u}(r)=\int_{\partial \Omega} \left ( \widetilde{u}_0(t)
\frac{\partial}{\partial n} \frac{e^{ik|r-t|}}{|r-t|} -
\frac{\partial}{\partial n}\widetilde{u}_0(t)
\frac{e^{ik|r-t|}}{|r-t|}\right ) dS(t),
\end{equation}
where function $\widetilde{u}_0$ is at least twice differentiable
in $\overline{\Omega '}$ (see \cite[formula 2.3']{majda} for
detailes) and therefore functions $\widetilde{u}_0(t)$ and
$\frac{\partial}{\partial n}\widetilde{u}_0(t)$ are bounded on
$\partial \Omega$.

>From (\ref{majaux2}) we obtain that exists such
$\widetilde{C}=\widetilde{C}(\gamma)$ such that for all $\theta
\in S^2$ and values of $k$ large enough
$$
|\widetilde{f}_\gamma(\theta)|<\widetilde{C}k,
$$
and due to (\ref{majaux1}) follows that the same bound (maybe with
another constant $\widetilde{C}_2$) is valid for $f(\theta)$.
Therefore, using optical theorem (see \cite{Kriegsmann} for proof
in case of impedance boundary condition )
$$
 \sigma_\gamma = \frac{4\pi}{k} Im [f_\gamma(\theta_0)],
$$
we obtain that
\begin{equation}\label{almlast}
\sigma_\gamma \leq 4 \pi \widetilde{C}_2.
\end{equation}
This ends the proof of the first part of the theorem \ref{maj}.

>From (\ref{uuogamma}) we  have for $k \rightarrow \infty$
\begin{equation}\label{majlem}
|f_\gamma(\theta)|^2=\frac{1}{4}\mathcal K(y^+(\theta))^{-1}\left
( \frac{\gamma - <\bsn(\theta) , \theta>} {\gamma + <\bsn(\theta),
\theta>} \right )^2+O(1
 / k), \quad \theta \neq \theta_0,
\end{equation}
where estimation $O(1/k)$ in (\ref{majlem}) is uniform on compact
subsets of $\{ \theta \in S^2 | \theta \neq \theta_0 \}$.

Using (\ref{almlast}) and the fact that density of the integral (\ref{rrr})
 is continuous and it turns to zero in the point
$\theta=\theta_0$, we obtain for $k \rightarrow \infty$
\begin{equation}\label{majlem2}
\begin{split}
& R_\gamma= \int_{S^2}(1-<\theta,\theta_0>) |f_\gamma(\theta)|^2 d
\sigma(\theta) =
\\ & \int_{S^2} (1-<\theta,\theta_0>) (4\mathcal K(y^+(\theta)))^{-1}\left (
\frac{\gamma - <\bsn(\theta) , \theta>} {\gamma + <\bsn(\theta),
\theta>} \right )^2 d\sigma (\theta)+o(1),
 \end{split}
\end{equation}

Let us construct a change of variables $\theta(x) : \mathcal I
\rightarrow S^2$. Let $\bsn(x)$ be an outward normal in the point
$y^+(x)=(x,g(x)) \in
\partial \Omega$, then we put
$\theta(x)=\theta_0-2<\bsn(x),\theta_0>\bsn$. It's easy to see that
$\theta(x) \in S^2$ and this map is one-to-one, since the obstacle
$\Omega$ is strictly convex and therefore the Gauss map $\bsn(x)$
is also one-to-one.

Let us introduce standard spherical coordinates
$(\cos(\widetilde{\theta}),\widetilde{\varphi}) \in [-1,1] \times
[0,2\pi)$ and we will calculate Jacobian
$D(\cos(\widetilde{\theta}),\widetilde{\varphi})/D(x_1,x_2)$,
where $(x_1,x_2)=x$ are orthonormal coordinates on $\mathcal I$.

 Note that
 $$
\theta(x)=\left ( \frac{2g'_{x_1}}{1+|\nabla g|^2},
\frac{2g'_{x_2}}{1+|\nabla g|^2}, \frac{|\nabla g|^2-1}{|\nabla
g|^2+1} \right)
 $$
$$
\begin{array}{c}
\cos(\widetilde{\theta})= 1-\frac{2}{|\nabla g|^2+1}, \\
\widetilde{\varphi} = \arctan (g'_{x_2}/g'_{x_1})
\end{array}
$$
$$
\frac{D(\cos(\widetilde{\theta}),\widetilde{\varphi})}{D(x_1,x_2)}=
\left |
\begin{array}{cc}
\cos(\widetilde{\theta})'_{x_1} & \widetilde{\varphi}'_{x_1} \\
\cos(\widetilde{\theta})'_{x_2} & \widetilde{\varphi}'_{x_2}
\end{array}
\right | =
$$
$$
\frac{4}{(1+|\nabla g|^2)(|\nabla g|^2)} \left |
\begin{array}{cc}
g'_{x_1}g''_{x_1 x_1} + g'_{x_2}g''_{x_1 x_2} & g'_{x_1}g''_{x_1 x_2} - g'_{x_2}g''_{x_1 x_1} \\
g'_{x_1}g''_{x_1 x_2} + g'_{x_2}g''_{x_2 x_2} & g'_{x_1}g''_{x_2
x_2} - g'_{x_2}g''_{x_1 x_2}
\end{array}
\right | =
$$
\begin{equation}\label{exlast0}
\frac{4 ((g''_{x_1 x_2})^2-g''_{x_1 x_1}g''_{x_2 x_2})}{1+|\nabla
g|^2}=-4\mathcal K(x_1,x_2).
\end{equation}
Note now that for $x \in \mathcal I$
\begin{equation}\label{exlast}
<\theta_0-\theta(x),\theta_0>=<2<\bsn(x),\theta_0>\bsn,\theta_0>=2<\bsn(x),\theta_0>^2=
\frac{2}{1+|\nabla u(x)|^2},
\end{equation}
Applying  (\ref{exlast}) and (\ref{exlast0}) for (\ref{majlem2})
we obtain (\ref{eqmaj}).
 Theorem \ref{maj} is proved.


\begin{thebibliography}{102}


\bibitem{majda} A.~Majda, High frequency Asymptotics for the Scattering matrix and
the inverse problem of Acoustical scattering, \emph{Comm. pure and
applied math.} vol. XXIX, 261--291, (1976)

\bibitem{majdaTay} A.~Majda, M.E.Taylor, The asymptotic behavior of the diffractive peak in classical scattering, \emph{Comm. pure and
applied math.} vol. XXX, 639--669, (1977)

\bibitem{Kriegsmann} Kriegsmann GA, Alternate proof of the optical theorem for
impenetrable targets, \emph{ J Acoust Soc Am.}, 2006
Jan;119(1):31-2.

\bibitem{vantychonovsamarsky} Tychonov, A. N., and Samarsky, A. A., Equations
of Mathematical Physics Pergamon, Oxford, 1963.

\bibitem{AlRamm} Ramm A. G., Scattering by Obstacles (Dordrecht:
Reidel), 1986

\bibitem{sph} A.I.Aleksenko, W. de Roeck, E.L.Lakshtanov, Resistance of the Sphere to a Flow of Quantum Particles, J.Phys.
A.Math.Gen, (39), pp. 4251-4255, 2005.

\bibitem{wl} W. De Roeck, E.L.Lakshtanov, Total cross section exceeds transport cross section for quantum scattering from hard bodies at low and high wave numbers
, J.Math.Phys, 48, 2007.

\bibitem{gauss}
Banchoff, T., Gaffney T., McCrory C., Cusps of the Gauss Map,
(1982) Research Notes in Mathematics 55, Pitman, London.

\bibitem{Principia} I. Newton,\, {\it Philosophiae naturalis principia mathematica}\,
1686.

\bibitem{BK}
G. Buttazzo, B. Kawohl,\, {\it On Newton's problem of minimal
resistance},\, Math. Intell. {\bf 15}, No.4, 7-12 (1993).


\bibitem{P1}
A.\,Yu. Plakhov. {\it Newton's problem of a body of minimal
aerodynamic resistance}. Dokl. Akad. Nauk {\bf 390},
N$^\text{o}$3, 314--317 (2003).

\bibitem{CL1}
M. Comte, T. Lachand-Robert,\, {\it Newton's problem of the body
of minimal resistance under a single-impact assumption}.\, Calc.
Var. Partial Differ. Equ. {\bf 12}, 173-211 (2001).
\end{thebibliography}
\end{document}